\documentclass[prl, amsmath, showpacs, preprintnumbers,
superscriptaddress, twocolumn, sort&compress]{revtex4}

\usepackage{graphicx}
\usepackage{mathptmx}
\usepackage[latin1]{inputenc}

\begin{document}
\title{Long-lived Feshbach molecules in a 3D optical lattice}
\author{G. Thalhammer}
\author{K. Winkler}
\author{F. Lang }
\author{S. Schmid}
\affiliation{Institut f\"ur Experimentalphysik, Universit\"at
  Innsbruck, 
  6020 Innsbruck, Austria}

\author{R. Grimm}%
\affiliation{Institut f\"ur Experimentalphysik, Universit\"at
  Innsbruck, 
  6020 Innsbruck, Austria}%
\affiliation{Institut f\"ur Quantenoptik und Quanteninformation,
  \"Osterreichische Akademie der Wissenschaften, 6020 Innsbruck,
  Austria}

\author{J. Hecker Denschlag}%
\affiliation{Institut f\"ur Experimentalphysik, Universit\"at
  Innsbruck, 
  6020 Innsbruck, Austria}

\date{\today}


\begin{abstract}
  We have created and trapped a pure sample of $^{87}$Rb$_2$ Feshbach
  molecules in a three-dimensional optical lattice.  Compared to
  previous experiments without a lattice we find dramatic improvements
  such as long lifetimes of up to 700\,ms and a near unit efficiency
  for converting tightly confined atom pairs into molecules. The
  lattice shields the trapped molecules from collisions and thus
  overcomes the problem of inelastic decay by vibrational quenching.
  Furthermore, we have developed a novel purification scheme that
  removes residual atoms, resulting in a lattice in which individual
  sites are either empty or filled with a single molecule in the
  vibrational ground state of the lattice.
\end{abstract}

\maketitle

Using magnetic Feshbach resonances \cite{feshbach} to create ultracold
diatomic molecules in their highest ro-vibrational state has become a
key to exciting developments and breakthroughs. Above all, it is the
highly versatile control of the interactions and binding energy of
these weakly bound molecules which opens the door to various
fundamental systems such as strongly-interacting quantum-degenerate
Fermi gases \cite{Che05} or Efimov few-body states \cite{Bra03}.

Feshbach molecules made of bosonic atoms behave in a strikingly
different way from Feshbach molecules made of fermionic atoms.  For
weakly bound dimers of fermionic atoms, vibrational quenching and
inelastic decay is strongly suppressed by a Pauli blocking effect in a
close encounter of two molecules \cite{Pet04}. This has been vital to
the experimental creation of molecular Bose-Einstein condensates (BEC)
and investigations of the crossover to a strongly interacting
fermionic superfluid \cite{Che05}. For dimers of bosonic atoms
\cite{Don02a,Her03,Xu03,Dur04}, however, progress has been hampered by
strong inelastic decay due to atom-molecule and molecule-molecule
collisions. Therefore the experiments have been focussed on the
transient regime, studying e.g. the collision and dissociation
dynamics \cite{Dur04,Xu03,Muk04,Dur04b,Hod05,Chi05}.

A three-dimensional optical lattice offers many interesting
opportunities for research on ultracold molecules. Lattice sites
occupied with exactly two atoms represent a perfectly controlled
quantum system which can be rigorously treated theoretically.  Matrix
elements for atom-molecule coupling are strongly enhanced with the
prospect of efficient atom-molecule conversion. Moreover, it is
expected that the lattice can isolate molecules from each other and
shield them from detrimental collisions so that a long-lived sample
can be created also with dimers of bosonic atoms. Recently, first
experiments with molecules in a lattice have studied photoassociation
\cite{Rom,Ryu} or demonstrated modifications of the binding energy of
tightly confined Feshbach molecules \cite{Stoeferle}.

In this Letter, we report on the creation of a pure sample of
ultracold Rb$_2$ Feshbach molecules trapped in a 3D optical lattice.
The observed long lifetimes of up to 700\,ms greatly exceed previous
values reported for dimers of bosonic atoms
\cite{Dur04,Xu03,Muk04,Dur04b}.  Further, we experimentally
investigate association and dissociation of the Feshbach molecules and
reach efficiencies of 95\% for converting pairs of atoms into
molecules. In brief, we adiabatically load a $^{87}$Rb BEC into the
vibrational ground state of the lattice. For our experimental
conditions, about 20\% of the condensate atoms are grouped in pairs of
two into the lattice sites. By ramping adiabatically over a magnetic
Feshbach resonance at 1007.4\,G we convert these pairs into molecules.
Another 20\% percent of atoms are located in triply and more highly
occupied lattice sites. After the Feshbach ramp, however, inelastic
collisions between the created molecules and atoms within the high
occupancy sites quickly remove these particles from the lattice.
Finally, the remaining 60\% of the condensate atoms are found in
singly occupied sites and are unaffected by the Feshbach ramp. Using a
novel resonant purification scheme we can remove these atoms from the
lattice, which results in a pure molecular sample with each molecule
being shielded from the others by the lattice potential.

The starting point for our experiments is an almost pure BEC of about
$6\times 10^5$ $^{87}$Rb atoms in the spin state $| F = 1, m_F = -1
\rangle$ \cite{Tha05}. It is transferred from a quadrupole Ioffe
configuration trap (QUIC) into a magnetic trap \cite{movBEC} with trap
frequencies $\omega_{x,y,z} = 2\pi\times (7, 19, 20)$ Hz, leading to a
peak density of the BEC of about $4\times 10^{13}$\,cm$^{-3}$.  Our 3D
lattice is cubic and consists of three retro-reflected
intensity-stabilized laser beams which propagate orthogonally to each
other. They are derived from a frequency-stable single-mode
Ti:Sapphire laser ($\approx$ 500 kHz linewidth) with a wavelength of
$\lambda = $ 830.44\,nm. For this wavelength, the laser is detuned by
about 100 GHz from the closest transition to an excited molecular
level, minimizing light induced losses as a precondition for long
molecular lifetimes. The laser beams are polarized perpendicularly to
each other and their frequencies differ by several tens of MHz to
avoid disturbing interference effects. The waists of all three beams
are about 160\,$\mu$m and the maximum obtainable power is about
130\,mW per beam, which results in calculated lattice depths of up to
40 recoil energies ($E_{r}= h^2 / 2m \lambda^{2}$, where $m$ is the
atomic mass of $^{87}$Rb and $h$ is Planck's constant). We have
verified the lattice depths by measuring the energy gap between bands
of the lattice \cite{denschlag}. The relative uncertainty of our
lattice depth is $\pm 15\%$.

After the BEC is adiabatically loaded into a 35 $E_r$ deep 3D optical
lattice within 100\,ms, we turn off the magnetic trap. By suddenly
reversing the bias magnetic field of a few G we flip the spins of our
atoms to the high field seeking state $| F = 1, m_F = +1 \rangle$ with
an efficiency higher than 99\% (see also \cite{Vol03}). This state
features the Feshbach resonance at 1007.4 G. Afterwards we ramp up a
homogeneous magnetic field in 3\,ms to about 1015\,G using the QUIC
quadrupole coils in Helmholtz configuration. The current through the
coils is actively stabilized to a relative accuracy of about
$10^{-4}$. The fast diabatic crossing of the Feshbach resonance has
basically no effect on the atoms in the lattice. If we slowly ramp in
5\,ms from 1015\,G to 1000\,G (crossing the Feshbach resonance at
1007\,G) molecules are adiabatically produced in the multiply occupied
lattice sites. If, however, we cross the Feshbach resonance very
quickly e.g. by simply switching off the magnetic field, less than
$10\%$ of the atoms are converted into molecules.  Note, that after
the first Feshbach ramp, we observe an immediate irretrievable loss of
20\% of the atoms. We attribute this loss to inelastic collisions
involving molecules for sites initially occupied by 3 or more atoms.
The remaining occupied sites contain each either a single atom or a
single molecule.

Atom numbers are measured with absorption imaging at low magnetic
fields ($\approx$ 2G) after release from the optical lattice and
11\,ms of ballistic expansion. In order to determine molecule numbers,
they are first dissociated into atoms by slowly ramping back across
the Feshbach resonance and then quickly switching off the magnetic
field.  We also use absorption imaging to map out the band occupation
of the lattice.  For this, the lattice is ramped down in 2\,ms and we
typically observe a momentum distribution which is fully contained in
a cube of width $2\,\hbar k$ corresponding to the first Brillouin zone
of the lattice \cite{greiner01}. This demonstrates that atoms and
molecules are in the vibrational ground state of the lattice sites.

In order to create a pure molecular sample, we have developed a novel
purification scheme to remove all atoms. We apply a combined microwave
and light pulse at a magnetic field of 1000\,G for 3\,ms. The
microwave drives the transition at a frequency of 9113\,MHz between
levels which correlate with $| F = 1, m_F = +1 \rangle$ and $|F=2, m_F
= +2 \rangle$. The light pulse drives the closed transition $| F = 2,
m_F = +2 \rangle \rightarrow | F = 3, m_F = +3 \rangle$. The optical
transition frequency is 1402\,MHz blue detuned compared to the
transition at zero magnetic field. After this pulse which heats the
atoms out of the lattice and an additional hold time ($\sim$20\,ms),
no more atoms can be detected. The direct effect of the microwave and
light field pulse on the molecules is negligible because the radiation
is off resonance.  As an indirect effect, however, we find that during
the first purification pulse we still lose about 40\% of the
molecules, probably due to inelastic collisions with the blown away
atoms. Further losses are not observed in subsequent purification
pulses. We end up with a pure molecular sample formed from about 10\%
of the initial atoms, which corresponds to 3$\times 10^4$ molecules.

\begin{figure}
  \includegraphics{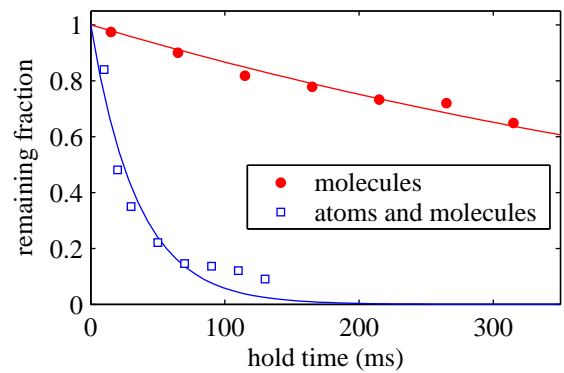}
  \caption{Decay of molecules in a 3D optical lattice with a potential
    depth of $(10\pm 2)\,E_r$. Shown is the remaining fraction of
    molecules in purified (filled circles) and unpurified (squares)
    samples as a function of hold time.  The continuous lines are
    exponential fits to the data indicating a lifetime of 700\,ms and
    35\,ms for purified and unpurified molecular samples,
    respectively.  In order to determine molecule numbers in the
    unpurified sample, purification was performed at the end of the
    hold time.  The hold time was limited to 400 ms due to the
    heating-up of the coils.  }
  \label{fig1}
\end{figure}

\begin{figure}
  \includegraphics{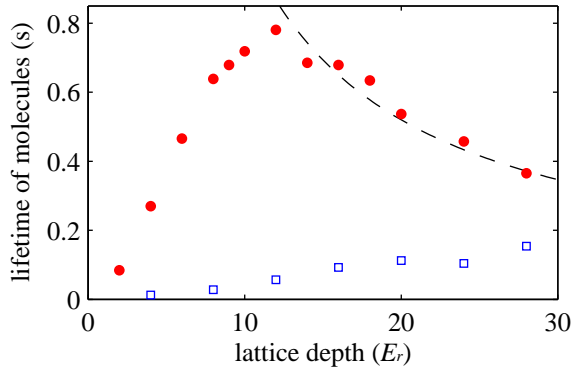}
  \caption{Molecular lifetimes for purified molecular samples
    (circles) and for unpurified samples (squares) as a function of
    the lattice depth. For this measurement the lattice depth was
    reduced from $35\,E_r$ to the given value after the creation of
    the molecules.  The dashed line is inversely proportional to the
    lattice depth.}
  \label{fig2}
\end{figure}

We have investigated the lifetimes of the Feshbach molecules in the
lattice under various conditions (see Figs.~\ref{fig1} and
\ref{fig2}). Figure~\ref{fig1} shows the decay of molecules at a
lattice depth of 10 $E_r$. The pure molecular sample exhibits a
remarkably long lifetime of 700\,ms. For the case of an unpurified
sample, where the atoms at singly occupied sites have not been
removed, the lifetime of the molecules is considerably reduced to
$\approx$35\,ms. This observation suggests that the molecular decay is
based on a process where an atom tunnels to a site occupied by a
molecule and inelastically collides with it. These inelastic
collisions can in principle also happen between two molecules.
However, compared to an atom, a molecule has a much lower tunnelling
rate, since it experiences twice the dipole potential and has twice
the mass than a single atom. Using simple scaling arguments, for a
molecule to have the same tunneling rate as an atom, the lattice light
intensity needs to be about four times smaller. This explains the
comparatively long lifetime of the purified molecular samples. We note
that if molecular decay is based on inelastic collisions, its time
dependence is intrinsically non-exponential. However, exact modelling
of the decay would be quite involved and requires precise knowledge of
atom/molecule distributions in the lattice. Since these distributions
are not known to us, we simply base our estimates for the molecular
lifetimes on an exponential decay law.

Figure~\ref{fig2} shows the measured lifetimes of the molecules for
various lattice depths. For sufficiently high lattice depths we
observe a lifetime for the purified molecular sample inversely
proportional to the lattice depth (see dashed line in
Fig.~\ref{fig2}). From this we conclude that above a lattice depth of
about $12\,E_r$ the tunnelling of the molecules is strongly suppressed
and the lifetime is limited by light induced losses.  Below this
value, decay is dominated by tunnelling and inelastic collisions. Thus
the molecular lifetime is maximized in a tradeoff between tunnelling
and light induced losses.  As already shown in Fig.~\ref{fig1}, the
presence of atoms considerably reduces the lifetime of the molecules,
even at larger lattice depths. In the limit of vanishing lattice
depths our experimental lifetimes decrease to values similar to those
observed in references \cite{Xu03,Muk04,Hod05,Dur04,Dur04b}.
Figure~\ref{fig2} clearly demonstrates that shielding of the molecules
against inelastic collisions grows with increasing lattice depth.

\begin{figure}
  \includegraphics{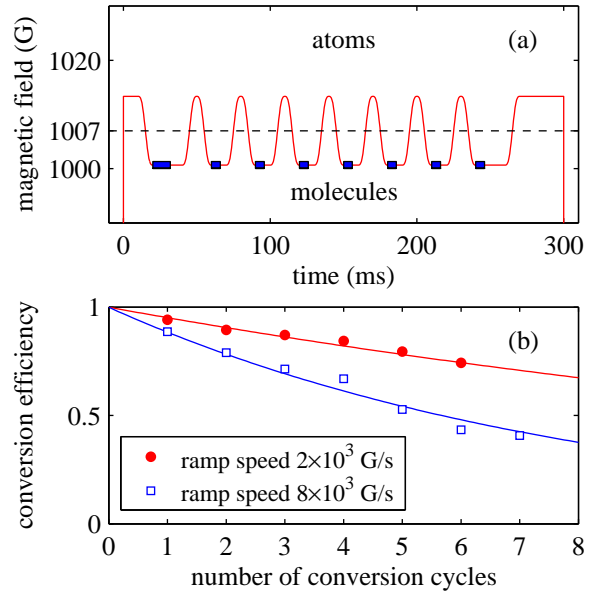}
  \caption{(a) Scheme for measurement of conversion efficiency, shown
    for 7 dissociation/association cycles. The shaded areas indicate
    the application of our purification procedure to remove atoms. The
    dashed line at 1007\,G shows the position of the Feshbach
    resonance.  (b) Conversion efficiency for given number of complete
    dissociation/association cycles for two different ramp speeds of
    the magnetic field.  We measure a conversion efficiency of 95\%
    per cycle for the slow ramp and 89\% per cycle for the fast ramp.
    The solid lines are described by exponential fitcurves as
    described in the text.  The lattice depth is 35 $E_r$.  }
  \label{fig3}
\end{figure}

We now investigate the dynamics for both association and dissociation
of a single Feshbach molecule in a lattice site during Feshbach
ramping. This fundamental system is of special interest since it can
be theoretically treated exactly and solved analytically \cite{Jul04}.
We prepare a purified sample of molecules at 1000 G in a lattice of 35
$E_r$ depth. We then ramp the magnetic field in a symmetric way across
the Feshbach resonance up to 1015 G and back (see Fig.~\ref{fig3}).
Afterwards, purification
is applied to remove atoms which have not recombined to form
molecules.  In a last step, the molecule number is measured. If
dissociation and association are not fully adiabatic in a conversion
cycle, a loss of molecules will result, e.g. during the association
ramp a pair of atoms might not be converted into a molecule or during
dissociation the molecule might break up into two atoms which, after
tunnelling, are located in separate sites.  For slow ramps we observe
small loss signals indicating high adiabaticity for the
dissociation/association cycle. In order to increase the loss signal,
to improve its accuracy and to check for consistency we repeated this
experiment with a higher number of cycles (see Fig.~\ref{fig3}). The
two data sets in (Fig.~\ref{fig3}b) correspond to two different ramp
speeds ($2\times 10^3\,\text{G/s}$, $8\times 10^3\,\text{G/s}$) and
can be described by the exponential functions $0.95^n$ and $0.89^n$,
respectively, where $n$ is the number of cycles.  Thus, for a slow
Feshbach ramp ($2\times 10^3\,\text{G/s}$) we observe an unprecedented
high efficiency of up to 95\% for the whole dissociation/association
cycle. For a faster ramp ($8\times 10^3\,\text{G/s}$) the efficiency
drops to 89\%.  We have taken care that light induced losses have been
corrected for in the data (Fig.~\ref{fig3}b). Our high conversion
efficiencies in the optical lattice are in strong contrast to the low
values of $ \sim $10\% observed previously in a $^{87}$Rb BEC
\cite{Dur04b} which were presumably limited by strong inelastic
collisions. In our deep lattice, however, inelastic collisions are
suppressed.

\begin{figure}
  \includegraphics{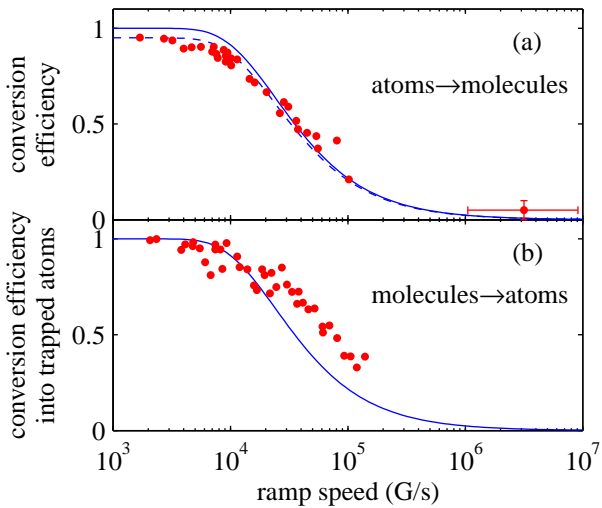}
  \caption{(a) Conversion efficiency of atoms (mostly pairs) into
    molecules as a function of the ramp speed. (b) A purified sample
    of molecules is dissociated into atom pairs at different ramp
    speeds.  We measure the number of atoms which are observed in the
    first Brillouin zone of the lattice after release, i.e., atoms
    that populate the lowest energy band of the lattice. The data is
    normalized to the atom number at the lowest ramp speeds. The
    continuous lines in (a) and (b) are calculations as described in
    the text.}
  \label{fig4}
\end{figure}

After having determined the efficiency for the full
dissociation/association cycle, we now study dissociation and
association individually. Figure~\ref{fig4}a shows the measured
conversion efficiency of atom pairs to molecules for different ramp
speeds. The atom pairs were prepared by creating a pure molecular
sample and then dissociating the molecules by slowly ($2 \times
10^3$G/s) ramping backward over the Feshbach resonance.  Then again
the magnetic field was swept across the Feshbach resonance at various
speeds and finally after switching off completely the magnetic field,
the remaining, non converted atoms are detected. The dashed line in
Fig.~\ref{fig4}a is based on a Landau-Zener expression without
adjustable parameters \cite{Jul04} and is given by
\begin{equation}
  p = 1-\exp\left(  -
    \frac{2\sqrt{6}\hbar}{m \ a^{3}_\text{ho}}%
    \left|\frac{a_\text{bg} \ \Delta B}{\dot{B}}\right|\right),
  \label{equ:formel}
\end{equation}
where $p$ is the probability of creating a molecule, $a_\text{bg} =
100.5\,a_0$ the background scattering length, $\Delta B =
0.21\,\text{G}$ \cite{Dur04b} the width of the Feshbach resonance,
$\dot B$ the ramp speed at the Feshbach resonance, and $a_\text{ho} =
\sqrt{{\hbar}/{m\omega}}$ the harmonic oscillator length. Using the
best estimate for our trapping frequency of $\omega = 2\pi \times (39
\pm 3)\,\text{kHz}$ (corresponding to a lattice depth of $35\pm
5\,E_r$) we get good agreement with our data. We note, that even for
the slowest ramp speeds the measured conversion efficiency never
reaches unity but levels off at 95\%, in agreement with the results in
Fig.~\ref{fig3}. This, however, does not exclude a true unit
conversion efficiency for atom pairs into molecules, because it is
possible that 5\% of the
atoms are not grouped in pairs, e.g. due to non-adiabaticity in
dissociation and tunnelling.  In order to facilitate the comparison of
the data distribution and theory, we have scaled the Landau-Zener
curve by a factor of 0.95 (dashed line). The maximum controllable ramp
speed ($\sim 10^5$ G/s, see Fig.~\ref{fig4}) is limited by the
performance of our current supply for the magnetic field coils. The
data point at $3\times 10^6$ G/s was obtained by simply switching off
the coil currents with an external switch.  The abrupt switching
induces eddy-currents which results in a less controlled ramp with a
large error margin. For fast switching we measured an conversion
efficiency of 5\%$\pm$5\%.

In Fig.~\ref{fig4}b we study the dissociation of a purified sample of
molecules. We determine the number of atoms which populate the lowest
band of the lattice after dissociation. At low ramp speeds Feshbach
molecules get adiabatically converted to pairs of atoms in the lattice
ground state. At higher speeds molecules are energetically lifted
above the molecule threshold and can decay into higher lattice bands
or into the continuum.  Assuming the reversibility of the Landau-Zener
transition, we use the same theory curve as in Fig.~\ref{fig4}a.  For
higher ramp speeds, we measure larger atom numbers than expected.
This is probably due to imperfections of our data analysis which can
overestimate the atom number in the lowest band by adding in some atoms
from higher bands.

To summarize, we have demonstrated that ultracold Feshbach molecules
can be created with high conversion efficiency in a 3D optical
lattice. After purification we observe long molecular lifetimes up to
700\,ms. These strong improvements over previous experiments open
promising perspectives for applications, e.g. in high resolution
molecular spectroscopy and quantum information processing in optical
lattices. They may also represent an important step in the creation of
a stable BEC of molecules in their vibrational ground state.

We appreciate the assistance of Matthias Theis and Carlo Sias. We
thank Thorsten Köhler and Todd Meyrath for fruitful discussions.  This
work was supported by the Austrian Science Fund (FWF) within SFB 15
(project part 17) and the European Union in the frame of the Cold
Molecules TMR Network under contract No.~HPRN-CT-2002-00290.



\begin{thebibliography}{99}
\bibitem{movBEC} After the initial production of the BEC in a
  quadrupole Ioffe configuration trap (QUIC) \cite{Tha05}, it is moved
  over a distance of $\approx$8 mm in order to be better centered on
  the QUIC quadrupole coils which are later used to generate the
  homogeneous magnetic field for Feshbach ramping. The relocation of
  the trap is done by adjusting the currents through the quadrupole
  and Ioffe coils and by applying additional magnetic field gradients.
\end{thebibliography}
\end{document}